\input harvmac
\noblackbox

%
%
\def\inbar{\,\vrule height1.5ex width.4pt depth0pt}
\def\IB{\relax{\rm I\kern-.18em B}}
\def\IC{\relax\hbox{$\inbar\kern-.3em{\rm C}$}}
\def\ID{\relax{\rm I\kern-.18em D}}
\def\IE{\relax{\rm I\kern-.18em E}}
\def\IF{\relax{\rm I\kern-.18em F}}
\def\IG{\relax\hbox{$\inbar\kern-.3em{\rm G}$}}
\def\IH{\relax{\rm I\kern-.18em H}}
\def\II{\relax{\rm I\kern-.18em I}}
\def\IK{\relax{\rm I\kern-.18em K}}
\def\IL{\relax{\rm I\kern-.18em L}}
\def\IM{\relax{\rm I\kern-.18em M}}
\def\IN{\relax{\rm I\kern-.18em N}}
\def\IO{\relax\hbox{$\inbar\kern-.3em{\rm O}$}}
\def\IP{\relax{\rm I\kern-.18em P}}
\def\IQ{\relax\hbox{$\inbar\kern-.3em{\rm Q}$}}
\def\IR{\relax{\rm I\kern-.18em R}}
\font\cmss=cmss10 \font\cmsss=cmss10 at 7pt
\def\IZ{\relax\ifmmode\mathchoice
{\hbox{\cmss Z\kern-.4em Z}}{\hbox{\cmss Z\kern-.4em Z}}
{\lower.9pt\hbox{\cmsss Z\kern-.4em Z}}
{\lower1.2pt\hbox{\cmsss Z\kern-.4em Z}}\else{\cmss Z\kern-.4em Z}\fi}

%
%

\def\({\left (}
\def\){\right )}

%
%

\def\NP{{\it Nucl. Phys.\ }}

\def\PL{{\it Phys. Lett.\ }}
\def\PR{{\it Phys. Rev.\ }}
\def\PRL{{\it Phys. Rev. Lett.\ }}

\def\Mod{{\it Mod. Phys. Lett.\ }}

%
%
\font\tiau=cmcsc10

%
%
\lref\jp{J. Polchinski, hep-th/9510017, \PRL {\bf 75} (1995) 4274.}
\lref\vafwit{C. Vafa and E. Witten, hep-th/9507050.}
\lref\sqms{S. Ferrara, J.A. Harvey, A. Strominger and C. Vafa, hep-th/9505162,
Phys. Lett. {\bf B361} (1995) 59.}
\lref\ms{J. Maldacena and A. Strominger, {\it Statistical 
Entropy of Four Dimensional Extremal Black Holes},
 Phys. Rev. Lett. {\bf 77} (1996)
428, hep-th/9603060.}
\lref\cliffordfd{ C. Johnson, R. Kuhri and R. Myers, {\it 
Entropy of 4D Extremal Black Holes}, Phys. Lett. {\bf B378} (1996) 78,
 hep-th/9603061.}
\lref\ascv{A. Strominger and C. Vafa, {\it On the Microscopic 
Origin of the Bekenstein-Hawking Entropy}, Phys. Lett. {\bf B379} (1996)
99,
 hep-th/9601029.}
\lref\cama{C. Callan and J. Maldacena, hep-th/9602043, \NP {\bf B472}
 (1996) 591.}
\lref\kol{R. Kallosh and B. Kol, hep-th/9602014, \PR {\bf D53} (1996) 5344; 
R. Dijkgraaf,
E. Verlinde and H. Verlinde, hep-th/9603126.}
\lref\tf{A. Tseytlin, hep-th/9601177, \Mod {\bf A11} (1996) 689.}
\lref\openstrom{A. Strominger, hep-th/9512059, \PL {\bf B383} (1996) 44.}
\lref\opentown{P. Townsend, hep-th/9512062, \PL {\bf B373} (1996) 68.}
\lref\jmphd{ J. Maldacena, hep-th/9607235.}
\lref\spin{ J. Breckenridge, R. Myers, A. Peet and C. Vafa, hep-th/9602065. }
\lref\wittenpol{ J. Polchinski and E. Witten, hep-th/9510169, 
\NP {\bf B460} (1996) 506.} 
\lref\dougauge{M. Douglas, hep-th/9604198.}
\lref\msu{ J. Maldacena and L. Susskind, hep-th/9604042,
\NP {\bf B475}(1996) 679.}
\lref\classfour{ M. Cvetic and D. Youm, hep-th/9512127; M. Cvetic
and A. Tseytlin, hep-th/9512031, \PR {\bf D53} (1996) 5619.}
\lref\gimon{E. Gimon and J. Polchinski, hep-th/9601038, 
\PR {\bf D54} (1996) 1667.}
\lref\fks{S. Ferrara, R. Kallosh and A. Strominger, hep-th/9508072,
\PR {\bf D52} (1995) 5412.}
\lref\as{A. Strominger, hep-th/9602111, \PL {\bf B383} (1996) 39.} 
\lref\fk{S. Ferrara and R. Kallosh,
hep-th/9602136, \PR {\bf D54} (1996) 1514; 
hep-th/9603090, \PR {\bf D54} (1996) 1525.}
\lref\dmw{M.J. Duff, R. Minasian and E. Witten, hep-th/9601036, \NP
{\bf 465} (1996) 413.}
\lref\maharana{J. Maharana and J.H. Schwarz, hep-th/9207016, \NP
{\bf 390} (1993) 3.}
\lref\seiwit{N. Seiberg and E. Witten, hep-th/9603003,
 \NP {\bf B471} (1996) 121.}
\lref\berk{M. Berkooz, R.G. Leigh, J. Polchinski, J.H. Schwarz, N. Seiberg
and E. Witten, hep-th/9605184.}

\lref\kt{I. Klebanov and A. Tseytlin, {\it 
 Intersecting M-branes as Four-Dimensional Black Holes},
 Nucl. Phys. {\bf B475} (1996)
179,
hep-th/9604166.}
\lref\vijay{V. Balasubramanian and F. Larsen, {\it 
On D-Branes and Black Holes in Four Dimensions}, hep-th/9604189.}
\lref\poltassi{J. Polchinski, {\it TASI Lectures on D-Branes},
hep-th/9611050 and references therein.}
\lref\bcwkm{K. Behrndt, G. Lopes Cardoso, B. de Wit, R. Kallosh,
D. Lust, T. Mohaupt, {\it Classical and Quantum N=2 Supersymmetric
Black Holes}, hep-th/9610105.}

\lref\klms{ D. Kaplan, D. Lowe, J. Maldacena and 
A. Strominger, {\it Microscopic Entropy of N=2 Extremal Black Holes},
 hep-th/9609204.}

\lref\vafaanom{C. Vafa and E. Witten, 
{\it A One-Loop Test Of String Duality},  Nucl. Phys. {\bf B447} (1995)
261, hep-th/9505053;  
M. Bershadsky, C. Vafa and V. Sadov, 
{\it D-Branes and Topological Field Theories}, Nucl. Phys. {\bf B463} 
(1996) 420, hep-th/9511222.}

\lref\andymini{S. Ferrara and R. Kallosh, 
{\it Supersymmetry and Attractors}, Phys. Rev. {\bf D54} (1996)
1514,
hep-th/9602136; A. Strominger,
{\it  Macroscopic Entropy of N=2 Extremal Black Holes}, 
Phys. Lett. {\bf B383} (1996) 39,
hep-th/9602111.}

\lref\bm{ K. Berhndt and T. Mohaupt, ``{Entropy of N=2 Black
Holes and their M-brane Description}'', hep-th/9611140.}

\lref\wittenfive{ E. Witten, ``{Five-Brane Effective Action in
M-theory}'', hep-th/9610234.}

\lref\wittensugra{ E. Witten, ``{String Theory Dynamics in Various 
Dimensions}'', Nucl. Phys. {\bf B443} (1995) 85, hep-th/9503124.}

%
%
\baselineskip 12pt
\Title{\vbox{\baselineskip12pt 
\line{\hfil  RU-96-103}
\line{\hfil \tt hep-th/9611163}}}
{\vbox{\hbox{\centerline{N=2 Extremal 
Black Holes and Intersecting Branes}}}}
\centerline{\tiau 
Juan M. Maldacena\foot{malda@physics.rutgers.edu} 
}
\vskip .1in
\vskip.1in
\centerline{{\it  Department of Physics and Astronomy }}
\centerline{\it Rutgers University}
\centerline{\it Piscataway, NJ 08855, USA}
\vskip .5cm
\centerline{\bf Abstract}
\noindent
Using a simple hypothesis about the degrees of freedom of  intersecting
branes we find a microscopic counting argument that reproduces  the entropy 
of a class of BPS black holes of type IIA string theory on
general Calabi Yau three folds.

\Date{}


Recently \ascv\ the Bekenstein-Hawking area law for the entropy
of black holes was given a statistical interpretation
by counting the number of microstates in string theory 
configurations involving D-branes \poltassi .
These results were first obtained in $N=4,8$ supergravity theories in
four and five dimensions  \ascv \ms \cliffordfd\ and  they were extended
to special cases of $N=2$ theories in \klms . 
In this paper we study black holes in more general $N=2$ theories
corresponding to type IIA string  theory compactified on Calabi-Yau three
folds (CY$_3$) which can also be viewed as M-theory on
CY$_3\times S_1$. 
In type IIA string theory on a generic Calabi Yau manifold 
there are $1+ h_{11}$  different electric  charges
corresponding to 
D-zerobrane charge  and to  D-twobrane charges associated 
to the  different
homology classes of the Calabi Yau, as well as their
magnetic duals: D-sixbrane charge and D-fourbrane charges. 
We focus on the limit of large
Calabi Yau size in which   instanton corrections to the prepotential 
are small.
We  consider here black holes that carry zero brane charge plus
arbitrary fourbrane charge $(q_0, p^a)$ but they  have zero two brane and
six brane charges $ p^0=q_a=0$. $(q_0, p^a)$ are integers measuring the
number of D-zerobranes and D-fourbranes. 
In this  limit entropy is given in terms of the intersection
numbers $C_{abc}$ of the Calabi Yau.

The classical horizon area entropy of supersymmetric black holes
depends only on the charges and is proportional to
the square of the value of the central charge at the horizon
(in four dimensions) \andymini .
The central charge of the $N=2$ supersymmetry algebra 
is a moduli dependent combination
of the charges mentioned above and, through the BPS condition,
is equal to the mass of the black hole (in Einstein metric). 
The scalar fields at the horizon take the values that
minimize the central charge, or mass in Einstein metric,
for fixed values of the black hole charges.
The calculation of  the entropy of a BPS 
charged black hole reduces to  minimizing the central charge of
the supersymmetry algebra 
 with respect to the value of
the scalar fields. These are the  values that the scalar fields
take at the horizon which are
determined by the charges, we replace back these values in 
the expression for the central charge, we square it 
 and then we get the entropy \andymini . 
All we need to know is the expression of the central charge in 
terms of the charges. This is encoded by the prepotential.
The prepotential has a leading term involving the intersection
numbers $C_{abc}$ then
some  subleading terms, one involving the second Chern class
$c_2$, another involving the Euler number and finally some
instanton corrections. In the limit of big Calabi-Yau manifolds 
only the leading term survives. 
Since the entropy is given by the
value of the central charge at the horizon we need
the Calabi Yau to be  big at the horizon. This in turn implies the 
following 
condition on the charges $ q^0 \gg p^A$. This condition
can be understood 
from the BPS mass formula since 
the zerobranes want to make the Calabi Yau big while the 
fourbranes want to make it smaller. It can be checked that
the solutions for the moduli in terms of the charges imply that
the sizes of the Calabi Yau are stabilized by the two opposite 
tendencies of the zerobranes and fourbranes \bcwkm . 
Using this method the classical entropy formula  was derived in 
\bcwkm\
\eqn\entropycl{
S = 2 \pi \sqrt{ (q_0 + W_{0e}p^e) C_{abc}p^a p^b p^c/6}
}
with 
$$W_{0e} = {1\over 24} \int_{CY} c_2 \wedge J_e $$
where $J_a $ is a basis of the cohomology group H$^2$(CY,{\bf Z})
which is related to the homology basis of four cycles used to define 
the charges in
the usual way.
Equation \entropycl\ corresponds to
 the limit where the size of the Calabi Yau at the horizon is 
large.
We have kept 
a small correction which is the shift in the zero
brane charge since it is the leading correction and has a simple
 explanation.
We have neglected a correction going like $(p/q)^{3/2}$ involving
the Euler character as well as possible exponential corrections
related 
to instanton contributions \bcwkm .

The D-brane counting can be done using the ideas considered in 
\kt \vijay  . Notice that in six dimensions three fourbranes will
generically all intersect each other at a point. 
That counting was based on the following assumption:

{\it If three fourbranes intersect in a supersymmetric fashion
 at a point there
exists bound states of zero branes living at the intersection
point with 4 bosonic and 4 fermionic 
 degrees of freedom and
any zerobrane charge}.

This can be translated into a more transparent statement in 
 M-theory language. Consider
M-theory  on CY$_3 \times S_1$.
The fourbranes are M-theory fivebranes which also wrapp along the 
circle and the zerobrane charge is the KK momentum along the $S_1$
(11$^{th}$ direction). Here the assumption mentioned above
corresponds to the statement that {\it there are 4 massless bosonic
and 4 fermionic modes moving along the common intersection line of the 
three fivebranes} \kt .  

Actually what is actually needed is that 
 the number of bosonic plus half the
number of fermionic states is six. So we could also have six bosonic 
states. It is not implied by supersymmetry that we should have the
same number of bosonic and fermionic states since the modes along 
branes do not break any additional supersymmetries. In other words
the theory along the intersection line has (4,0) supersymmetry and
the modes we are considering are left moving, so they do not need 
to have the same number of bosonic and fermionic states. 
It actually seems natural that one could have six bosonic states
related to the position of the intersection point in the six
dimensional
manifold. 

I do not know of a  direct proof of the above statements,
but there is some evidence from U-duality that this should be the case,
since the configurations in \kt ,\vijay\ are related to  a case where
the large number of degrees of freedom is easier to argue for \ms .
This hypothesis is  very simple and leads to a nice
geometrical picture for the origin of the entropy. It has been proposed
in \kt\ that these modes should be related to a twobrane connecting
the three fivebranes through a Y junction, which would be extra
massless 
states
appearing when three fivebranes intersect. 
It seems possible that this hypothesis  can be proved  using the 
analysis in \wittenfive .

On a general Calabi Yau manifold the four branes will intersect themselves
generically at $ N_{ int} = C_{abc}p^ap^bp^c/6 $ points. The factor of 6 is
a symmetry factor. 
The zero brane bound state can have charges $q_0^i  = 1, 2, 3,...$.
The counting of states is like the counting for a one dimensional gas,
a fact familiar from the type IIA to 11 dimensional supergravity
correspondence \wittensugra . From the M-theory point of view we
have a gas of massless particles propagating along 11$^{th}$ dimension
with total momentum $P = q_0/R_{11}$. There are different species of
particles giving a total central charge
$c_T = N_B + {1\over 2 } N_F = 6 N_{int}$ which gives the entropy
%
\eqn\entrocount{
S = 2 \pi \sqrt{ q_0  N_{int} }  
}
Inserting the value $N_{int} =  C_{abc}p^ap^bp^c/6 $
we
get a formula similar to \entropycl .
In order to include the shift in the zero brane charge we should
remember
that a four brane
carries an anomalous zero brane charge equal to\foot{This correction
was computed up to a sign \vafaanom\ 
so we do not check the sign of the shift.} \vafaanom
\eqn\anomalous{
-{1\over 24 } \int_{v^4} R \wedge R = -{1\over 24 } \int_{v^4} c_2  
}
Integrating the second Chern class over all the four branes gives
\eqn\secchern{
\delta q_0 = - {1\over 24 } \int_{p^a v^4_a } c_2  =- p^a \int_{CY} 
c_2 \wedge J_a = - W_{oa}p^a 
}
This implies that the real number of zero branes that 
we are free to specify is  $q_{eff} = q_0 -\delta q_0$.
Replacing $ q_0 \rightarrow q_{eff} $ we get precisely 
formula \entropycl , including the shift in the charge.

In conclusion a simple geometrical picture was given for the 
entropy of black holes in general Calabi-Yau compactifications 
based on the simple hypothesis of \kt \vijay .
It would be nice to extend the micoscopic counting arguments
to cases where more charges are non-zero, as well as  understanding
more precisely the role of instanton corrections to the entropy 
formula. I would also be nice to give a proof of the assumptions
in \kt , \vijay .

{\bf Acknowledgments}

I would like to thank A. Strominger, R. Kallosh  and
M. Douglas for discussions.
This  research is supported in part by
 DOE grant DE-FG02-96ER40559. After this work was completed we
received a preprint \bm\ where the same picture was proposed.

\listrefs
\bye